\documentclass{article}

\usepackage{PRIMEarxiv}
\usepackage[utf8]{inputenc} 
\usepackage[T1]{fontenc}    
\usepackage{url}            
\usepackage{booktabs}       
\usepackage{amsfonts}       
\usepackage{nicefrac}       
\usepackage{microtype}      
\usepackage{lipsum}
\usepackage{fancyhdr}       
\usepackage{graphicx}       
\graphicspath{{media/}}     
\usepackage{subcaption}
\usepackage{cite}
\usepackage{amsmath,amssymb}
\usepackage{longtable}
\usepackage{algorithmic}
\usepackage{textcomp}
\usepackage{soul}
\usepackage{tabularx}
\usepackage{multirow}
\usepackage{booktabs}
\newcommand{\mycomment}[1]{}
\usepackage{multicol,lipsum}
\usepackage{comment}
\usepackage[dvipsnames]{xcolor}
\def\BibTeX{{\rm B\kern-.05em{\sc i\kern-.025em b}\kern-.08em
    T\kern-.1667em\lower.7ex\hbox{E}\kern-.125emX}}

\usepackage{algorithmic}
\usepackage{graphicx}
\usepackage{textcomp}
\newcommand{\etal}{\textit{et al.}}
\def\BibTeX{{\rm B\kern-.05em{\sc i\kern-.025em b}\kern-.08em
    T\kern-.1667em\lower.7ex\hbox{E}\kern-.125emX}}

\title{MedSegNet10: A Publicly Accessible Network Repository for Split Federated Medical Image Segmentation
\thanks{*Chamani Shiranthika and Zahra Hafezi Kafshgari contributed equally to this work.}
\thanks{\textit{\underline{Citation}}: 
\textbf{Chamani Shiranthika, Zahra Hafezi Kafshgari, Hadi Hadizadeh, Parvaneh Saeedi. MedSegNet10: A Publicly Accessible Network Repository for Split Federated Medical Image Segmentation. Pages:1-20, DOI:000000/11111.}} 
}

\author{
  Chamani Shiranthika, Zahra Hafezi Kafshgari, Hadi Hadizadeh, Parvaneh Saeedi \\
  School of Engineering Science \\ 
  Simon Fraser University \\ 
  Burnaby, BC, Canada \\
  \texttt{\{csj5, zahra\_hafezi, hadi\_hadizadeh, psaeedi\}@sfu.ca} 
}

\begin{document}
\maketitle
\begin{abstract}
Machine Learning (ML) and Deep Learning (DL) have shown significant promise in healthcare, particularly in medical image segmentation, which is crucial for accurate disease diagnosis and treatment planning. Despite their potential, challenges such as data privacy concerns, limited annotated data, and inadequate training data persist. Decentralized learning approaches such as federated learning (FL), split learning (SL), and split federated learning (SplitFed/SFL) address these issues effectively. This paper introduces "MedSegNet10," a publicly accessible repository designed for medical image segmentation using split-federated learning. MedSegNet10 provides a collection of pre-trained neural network architectures optimized for various medical image types, including microscopic images of human blastocysts, dermatoscopic images of skin lesions, and endoscopic images of lesions, polyps, and ulcers, with applications extending beyond these examples. By leveraging SplitFed’s benefits, MedSegNet10 allows collaborative training on privately stored, horizontally split data, ensuring privacy and integrity. This repository supports researchers, practitioners, trainees, and data scientists, aiming to advance medical image segmentation while maintaining patient data privacy. The repository is available at: \url{https://vault.sfu.ca/index.php/s/ryhf6t12O0sobuX} (password upon request to the authors).
\end{abstract}


\section{Introduction}
\label{sec:introduction}
Medical image segmentation plays a crucial role in the healthcare sector, serving as a foundational task for precise diagnosis, treatment planning, and patient care~\cite{image_segmenttaion_1,image_segmenttaion_2}. The advent of decentralized training approaches, such as federated learning (FL)\cite{McMahan_2017}, split learning (SL)\cite{Gupta_2018}, and split federated learning (SplitFed learning/SFL)\cite{Thapa_2022_AAAI} has opened up innovative avenues for collaborative, privacy-preserving, and resource-constrained model training that are distributed across hospitals, research institutes, and healthcare related organizations. 
Despite the considerable potential of decentralized learning approaches, their implementation for medical image segmentation faces notable challenges. Among many challenges, one notable obstacle is the limited availability of well-designed and rigorously evaluated DL networks. 

In this paper, we introduce the first documented repository of SplitFed networks tailored specifically for medical image segmentation tasks. Aligned with the recent evolution of semantic segmentation networks, we have curated the top-performing models based on their reported effectiveness in medical segmentation applications~\cite{semantic_ref_1,semantic_ref_2}. We selected ten prominent segmentation models in this field, including UNet\cite{ronneberger_2015_UNet}, SegNet\cite{badrinarayanan_2017_Segnet}, DeepLabV3\cite{chen_2017_Deeplabv3}, DeepLabV3+\cite{chen_2018_Deeplabv3+}, RefineNet\cite{lin_2017_refinenet}, CGNet\cite{wu_2020_cgnet}, SUNet\cite{SU_Net_Yi_2020}, DUCK-Net\cite{dumitru2023_ducknet}, Attention UNet\cite{oktay_2018_attentionunet}, and Swin-UNet\cite{cao_2022swin}. We trained and tested the ten networks using three datasets: Human Against Machine with 10,000 training images (HAM10K) \cite{tschandl_2018}, KVASIR-SEG \cite{jha2020_kvasir}, and our proprietary Blastocysts dataset \cite{lockhart_2019}.

Specifically, we designed and implemented SplitFed versions of the ten selected models and provide them in a unified repository, named MedSegNet10. This repository serves as a comprehensive guide and a significant contribution to the field of medical image segmentation. MedSegNet10 provides a versatile platform for both novice and experienced practitioners, providing a valuable toolkit for medical image segmentation within the novel SplitFed learning paradigm. Additionally, it grants researchers consolidated access to relevant networks, enabling comprehensive comparative analyses of architectural differences among models. To the best of our knowledge, we are the first to offer such a repository.

The structure of this paper is as follows: Section~\ref{Sec2} provides an overview of SplitFed fundamentals and the current state-of-the-art in SplitFed for medical image segmentation, including its integration with emerging transformer architectures. It also discusses publicly accessible federated network repositories and offers insights into semantic segmentation and the architecture of our implemented split models, followed by a discussion on our split point selection decisions. Section~\ref{Sec3} details the experiments, results, and evaluations. Section \ref{future_works} highlights the potential future directions. This paper concludes by presenting the conclusions in Section~\ref{Conclusions}.

\section{Related works}
\label{Sec2}
This section reviews related work on SplitFed learning, publicly available federated repositories, and semantic segmentation, including notable semantic segmentation networks. Also, it offers insights into our decisions on split point selection for designing our split networks.

\subsection{Split Federated Learning (SplitFed)}
SplitFed is the combination of FL and SL\cite{Thapa_2022_AAAI}. In FL, a central server collaboratively builds a global model using data from multiple devices or clients without accessing sensitive local information~\cite{liu2020systematic, mcmahan2017communication}. During collaborative training, clients are tasked with training the entire model on their respective local platforms. Consequently, some clients encounter challenges, particularly in comparison to local training, stemming from limited computational capabilities or resource-constrained issues. SL overcomes this issue by splitting and residing the model parts at different locations, devices, or clients~\cite{vepakomma2018split}. While federated structures protect client data privacy by securely storing raw information on local devices, splitting the model further alleviates potential computational burdens on clients.  Therefore, SplitFed learning is particularly advantageous in scenarios where data cannot be centralized for any reason. It strikes a balance between collaborative model training with limited computational resources, and preserving the  privacy of individual user data. 

\begin{figure}[tb]
\centerline{\includegraphics[scale = 0.40]{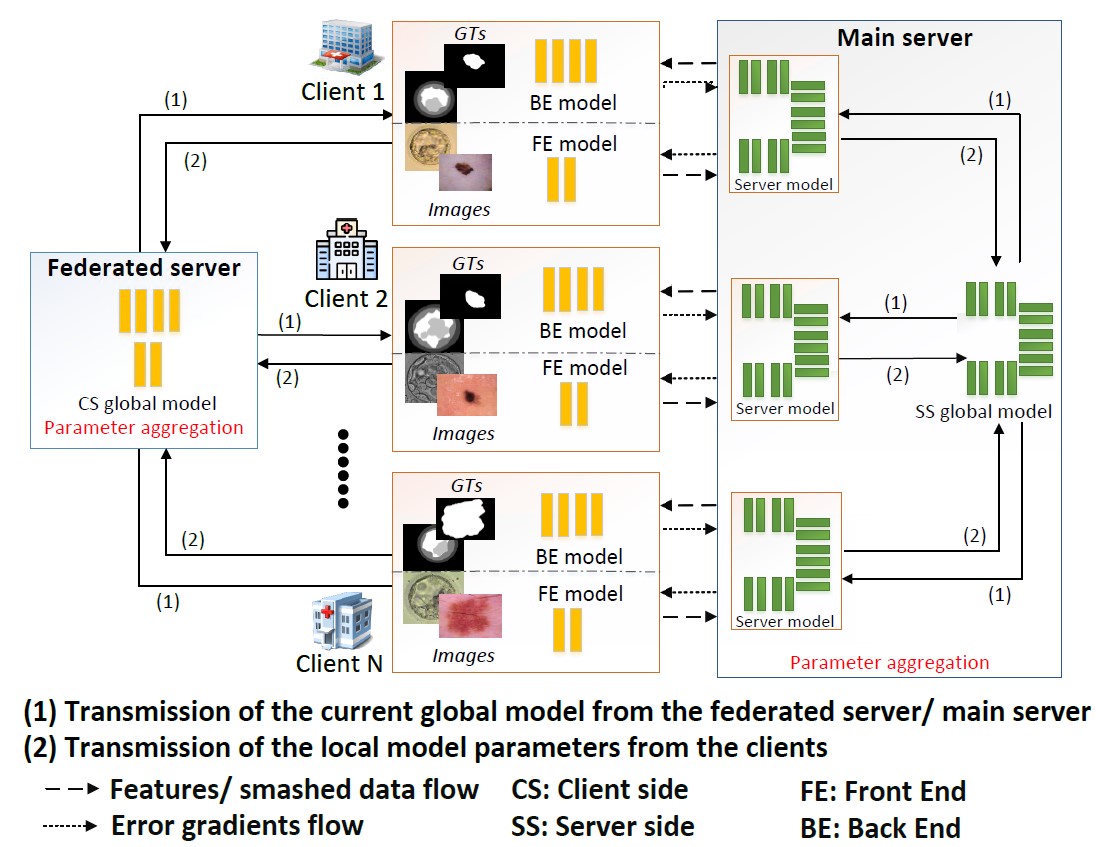}}
\caption{SplitFed architecture with multiple clients\cite{Thapa_2022_AAAI}, where each client uses the split UNet as its model\cite{Shiranthika_2023}.}
\label{fig:SplitFed}
\end{figure}

Fig.~\ref{fig:SplitFed} illustrates our SplitFed architecture, considering each client's network backbone as the UNet. In this U-shaped architecture, each client has two split points: the front-end (FE) and back-end (BE) sub-models, with the server sub-model in between. The FE sub-model exclusively accesses the images, while the BE sub-model solely interacts with the ground truth (GT) data. Both the FE and BE sub-models constitute a small portion of the UNet architecture and are located on the client side, while the server sub-model, containing the majority of the layers, handles the main computational burden of training. This splitting process was consistently applied across all ten selected models. 

SplitFed can be applied to various models, offering several advantages. The authors in \cite{Thapa_2022_AAAI} first highlighted SplitFed's effectiveness in maintaining data privacy and reducing communication overhead. The research in \cite{zhang2022splitavg} addressed data heterogeneity and improved model robustness using SplitFed, while \cite{yang2023dynamic} improved security through homomorphic encryption in a U-Shaped SplitFed network. Additionally, \cite{kafshgari2023quality} improved segmentation performance with imperfect labels, and \cite{kafshgari2023smart} tackled communication noise to ensure efficiency and accuracy.

Recently, transformers have emerged as a leading architecture in many DL applications. They excel at capturing long-range dependencies and complex patterns, which enhances the model's ability to learn intricate features from medical images. Their scalability and parallelism are particularly beneficial for federated learning scenarios, where efficient communication and computation are crucial. Integrating transformers into SplitFed networks has shown potential for improving performance in handling diverse, high-dimensional data while maintaining robustness and adaptability. Recent efforts, such as extending vision transformers (ViT) \cite{dosovitskiy2020image} in SplitFed \cite{baek2022visual, oh2022differentially}, aim to enhance performance, communication efficiency, and data privacy. Research in \cite{qu2022rethinking} introduced ViT-FL to address data heterogeneity in federated learning, demonstrating that transformers outperform conventional CNNs by providing better generalization in non-IID settings and achieving faster convergence.
Research in \cite{park2021federated} and \cite{Park_2023} applied SplitFed ViTs for diagnosing coronavirus disease (COVID-19) and chest radiographs (CXR), respectively. Also, FeSViBS \cite{almalik2023fesvibs} utilized ViTs on medical image datasets, including HAM10K \cite{tschandl_2018}, BloodMNIST \cite{acevedo2020dataset}, and Fed-ISIC2019 \cite{ogier2022flamby}. These studies collectively highlight the effectiveness of SplitFed transformers, demonstrating their significant performance improvements in medical imaging tasks.

\subsection{Publicly Available Federated Repositories}
Various public repositories for federated learning (FL), particularly those with annotated medical images \cite{Shiranthika_2023}, alongside numerous federated data analysis programs \cite{carter2016vipar, gazula2020coinstac, marcus2007extensible}, are available for public access. Notable federated network repositories include TensorFlow Federated \cite{TF}, Google Federated Research \cite{GFR}, LEAF \cite{LEAF}, Fedmint \cite{fedmint}, Federated Scope \cite{federatedscope}, FedLab \cite{zeng2021fedlab}, FATE \cite{FATE}, FedIIC \cite{Wu2023FedIIC}, QuickSQL \cite{Quicksql}, GraphQL-Mesh \cite{graphql}, MedAugment \cite{MedAugment}, OpenMined's PyGrid \cite{PyGrid}, FedCT \cite{FedCT}, and COALA \cite{zhuangcoala_2024}. Among these, FATE, MedAugment, OpenMined's PyGrid, and FedCT are specifically designed for medical imaging tasks.

Despite the availability of these resources, our survey reveals a gap in the current literature: the absence of a reusable SplitFed network repository that enables comprehensive comparisons and benchmarks. This gap highlights an opportunity for future research and development to create a robust and adaptable split network repository that supports collaborative networks. Our focus is on addressing this need by facilitating and streamlining the training processes for SplitFed networks in the healthcare domain.

\subsection{Image Segmentation Models}
Image segmentation is essential in healthcare for the accurate identification and extraction of regions within medical images, which is critical for disease detection, diagnosis, and treatment planning \cite{guo2018review}. By enhancing the capability of algorithms to analyze complex medical data, it plays a vital role in the interpretation of medical images. Deep Learning applications, such as SplitFed, further emphasize the significance of image segmentation by managing large-scale data across institutions while preserving privacy, thereby improving accuracy and efficiency. Semantic segmentation takes this further by classifying each pixel within identified regions, such as tumors or organs, providing detailed representations crucial for precise diagnosis and treatment. This section highlights MedSegNet10’s contributions to advancing semantic segmentation in medical imaging, addressing its complexities, and enhancing image content understanding.

Figure~\ref{fig:Timeline} illustrates a timeline of prominent semantic segmentation networks that have emerged over the past decades, with their full names and corresponding references listed in the order shown in the figure. These include:

\begin{figure*}[tb] 
\centerline{\includegraphics[scale = 0.45]{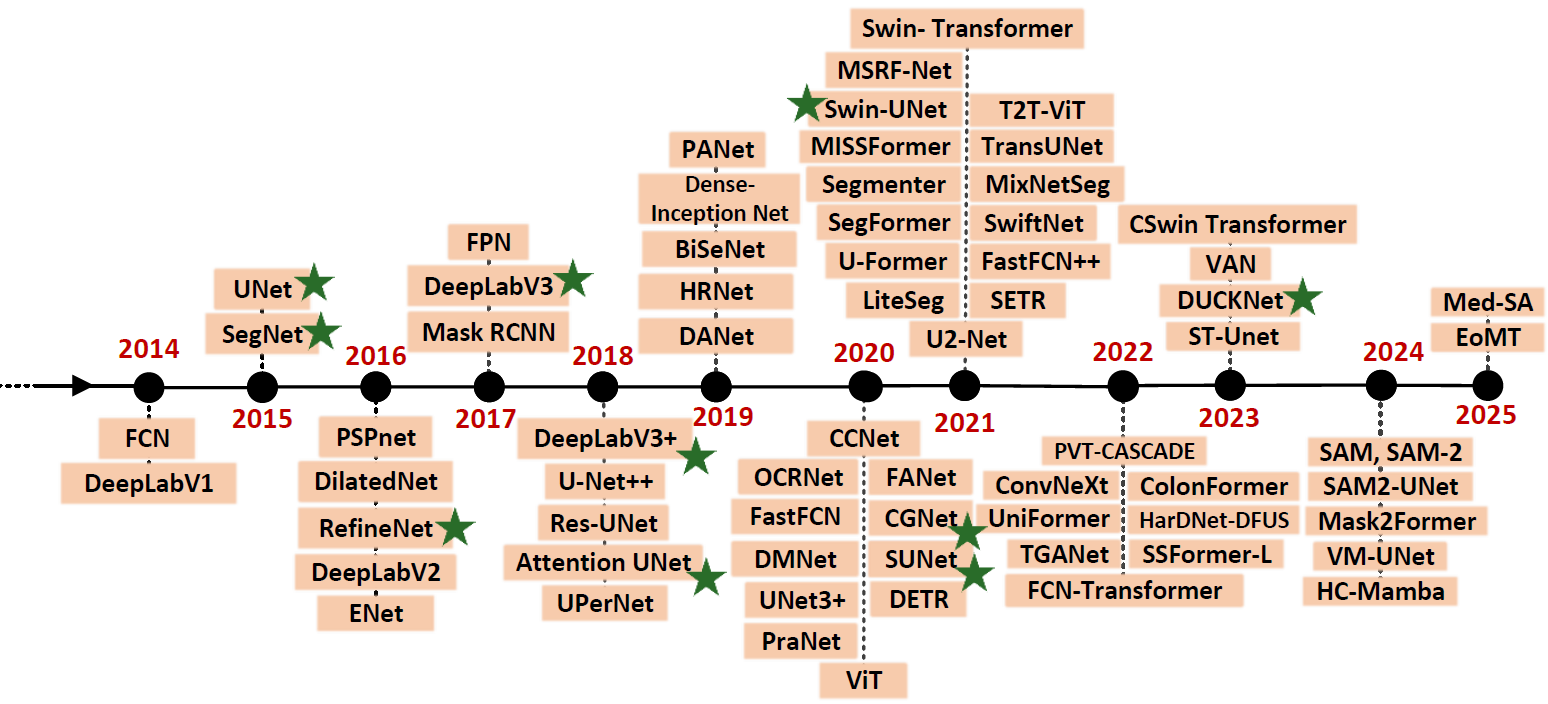}}
\caption{{\small{Semantic segmentation networks' development timeline. Networks marked with green stars are those included in our work.}}}
\label{fig:Timeline}
\end{figure*}

\begin{enumerate}
\item FCN\cite{long_2015_FCN}:  Fully Connected Network
\item {\color{ForestGreen}UNet~\cite{ronneberger_2015_UNet}: UNetwork}
\item {\color{ForestGreen}SegNet\cite{badrinarayanan_2017_Segnet}: Segmentation Network}
\item PSPnet\cite{zhao2017pyramid}: Pyramid Scene Parsing Network
\item ENet\cite{paszke2016enet}: Efficient Neural Network
\item {\color{ForestGreen}RefineNet: Refining Segmentation based Network}
\item {\color{ForestGreen}DeepLab\cite{chen_2014_Deeplab_V1,chen_2017_Deeplabv2,chen_2017_Deeplabv3,chen_2018_Deeplabv3+}: Deep Labelling for Semantic Image Segmentation}
\item {\color{ForestGreen}Attention UNet: Attention based UNet}
\item FPN\cite{lin2017_FPN}: Feature Pyramid Network
\item Mask RCNN\cite{he2017_maskRcnn}: Mask Region-based Convolutional Neural Network
\item PANet\cite{wang2019_panet}: Path Aggregation Network
\item BiseNet\cite{yu2018_bisenet}: Bilateral Segmentation Network
\item HRNet\cite{HRNet_2019}: High-Resolution Network
\item OCRNet\cite{yuan2020_OCRNet}: Object-Contextual Representations for Semantic Segmentation
\item DANet\cite{fu2019_DAnet}: Dual Attention Network
\item CCNet\cite{huang2019_ccnet}: Criss-Cross Attention Network
\item SETR\cite{SETR_heo2021}: Spatially-Enhanced Transformer
\item UPerNet\cite{xiao2018unified}: Unified Perceptual Parsing Network
\item FastFCN\cite{wu2019fastfcn}: Fast Fully Convolutional Network
\item {\color{ForestGreen}SUNet\cite{SU_Net_Yi_2020}: Strong UNet}
\item FANet\cite{singha2020_fanet}: Feature Aggregation Network
\item DMNet\cite{fang2020dmnet}: Dense Multi-scale Network
\item {\color{ForestGreen}CGNet\cite{wu_2020_cgnet}: Context-Guided Network}
\item DETR\cite{DETR_carion2020}: DEtection TRansformer
\item PraNet~\cite{fan2020pranet}: Parallel Reverse Attention Network
\item ViT~\cite{dosovitskiy2020image}: Vision Transformer
\item {\color{ForestGreen}Swin-UNet: Swin Transformer based UNet}
\item MSRF-Net~\cite{srivastava2021msrf}: Multi-Scale Residual Fusion Network
\item T2T-ViT\cite{yuan2021tokens}: Token-to-Token Vision Transformer
\item VAN\cite{guo2023visual}: Visual Attention Network
\item CSwin Transformer\cite{yang2023cswin}: Cross-Stage win transformer
\item {\color{ForestGreen}DUCK-Net\cite{dumitru2023using}: Deep Understanding Convolutional Kernel Network}
\item ST-UNet\cite{zhang2023st}: Spatiotemporal UNet
\item SAM\cite{kirillov2023segment}: Segment Anything 
\item VM‑UNet\cite{ruan2024vm}: Vision Mamba UNet
\item HC-Mamba\cite{xu2024hc}: Hybrid-convolution version of Vision Mamba
\item EoMT\cite{kerssies2025your}: Encoder-only Mask Transformer 
\item Med-SA\cite{wu2025medical}: Medical SAM Adapter
\end{enumerate}

We implemented both the Split and SplitFed versions of the networks marked with a green star in Fig.~\ref{fig:Timeline}. The following sections provide detailed descriptions of these implemented networks.

\subsubsection{UNet}
\begin{figure}[t]
    \centering
    \begin{subfigure}[b]{0.48\textwidth}
        \centering
        \includegraphics[width=\textwidth]{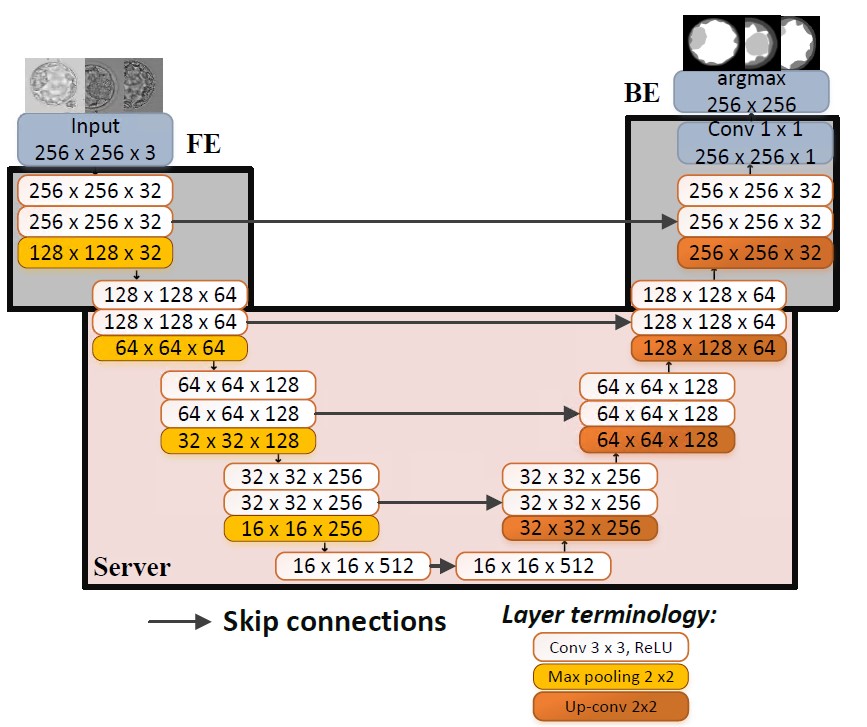}
        \caption{Split UNet architecture. Ash-colored segments are the layers on the client side, and pink-colored segments are the layers on the server side. Note that this naming convention is applied to all other network figures in this paper.}
        \label{fig:Unet}
    \end{subfigure}
    \hfill
    \begin{subfigure}[b]{0.48\textwidth}
        \centering
        \includegraphics[width=\textwidth]{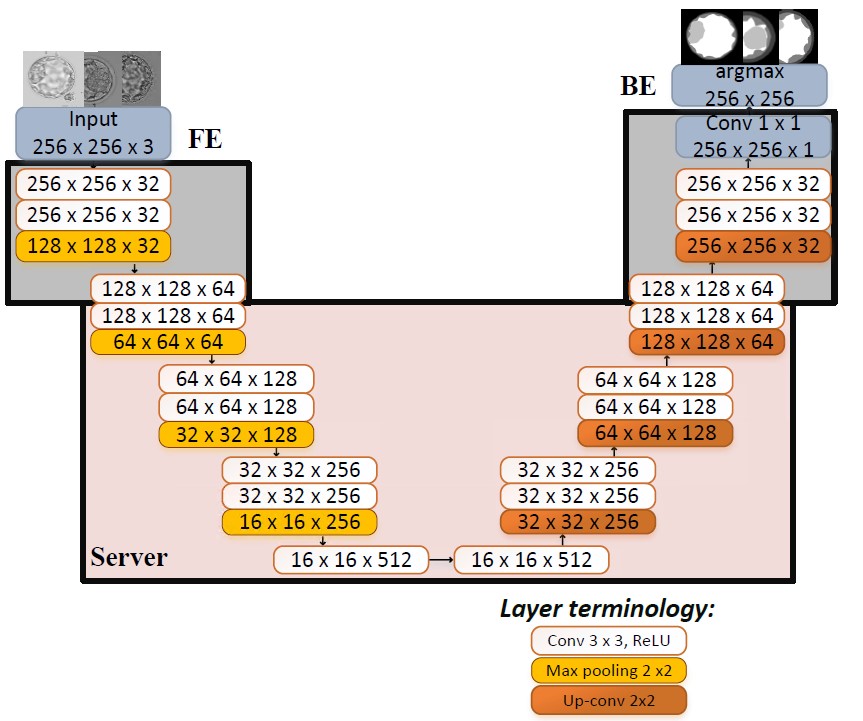}
        \caption{Split SegNet architecture. A similar architecture to UNet without the skip connections.}
        \label{fig:Segnet}
    \end{subfigure}
    \caption{Split UNet and SegNet architectures.}
    \label{fig:combined_unet_segnet}
\end{figure}

UNet can be considered as the most well-known architecture specially designed for image segmentation tasks\cite{ronneberger_2015_UNet}. It embodies the encoder-decoder architecture with skip connections. UNet employs a process of gradually upsampling the features extracted by its encoder using transpose convolutions. By incorporating skip connections, UNet reaches greater depths than many other state-of-the-art (SOTA) networks, resulting in significantly higher-quality outputs. Fig.~\ref{fig:Unet} shows the split version of the UNet used in this work. 

\subsubsection{SegNet} 
The SegNet architecture shares significant similarities with UNet as it adheres to an encoder-decoder framework\cite{badrinarayanan_2017_Segnet}. One of the distinguishing characteristics of SegNet from UNet is the absence of skip connections. Furthermore, SegNet diverges from the conventional upsampling operation found in many encoder-decoder architectures. Instead of traditional upsampling techniques, SegNet employs an approach referred to as ``max unpooling''. This technique mitigates the need to learn how to upscale both the final output score and the feature maps from earlier layers. In a conventional encoder-decoder structure, such upscaling operations typically demand substantial learning. By integrating max unpooling, SegNet optimally addresses this challenge. Fig.~\ref{fig:Segnet} depicts the split version of the SegNet used in our work.

\subsubsection{DeepLab}
DeepLab was developed by Google\cite{chen_2014_Deeplab_V1} and has been evolved to several versions including DeepLab\cite{chen_2014_Deeplab_V1}, DeepLabV1\cite{chen_2014_Deeplab_V1}, DeepLabV2\cite{chen_2017_Deeplabv2}, DeepLabV3\cite{chen_2017_Deeplabv3} and  DeepLabV3+~\cite{chen_2018_Deeplabv3+}. 

\begin{figure}[t]
    \centering
    \begin{subfigure}[b]{0.48\textwidth}
        \centering
        \includegraphics[width=\textwidth]{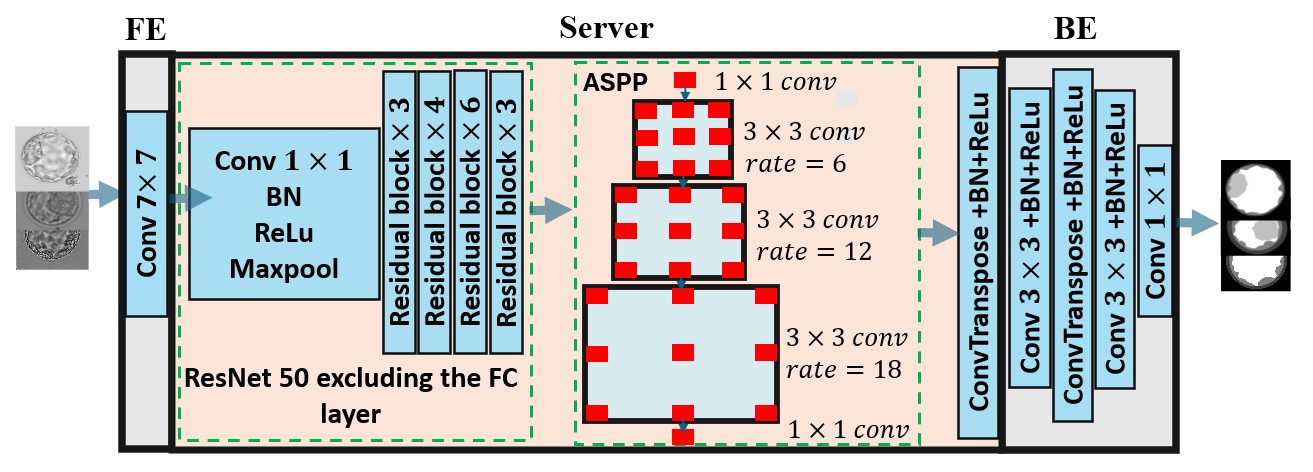}
        \caption{Split DeepLabV3 with ResNet50 as the backbone.}
        \label{fig:DeepLabV3}
    \end{subfigure}
    \hfill
    \begin{subfigure}[b]{0.48\textwidth}
        \centering
        \includegraphics[width=\textwidth]{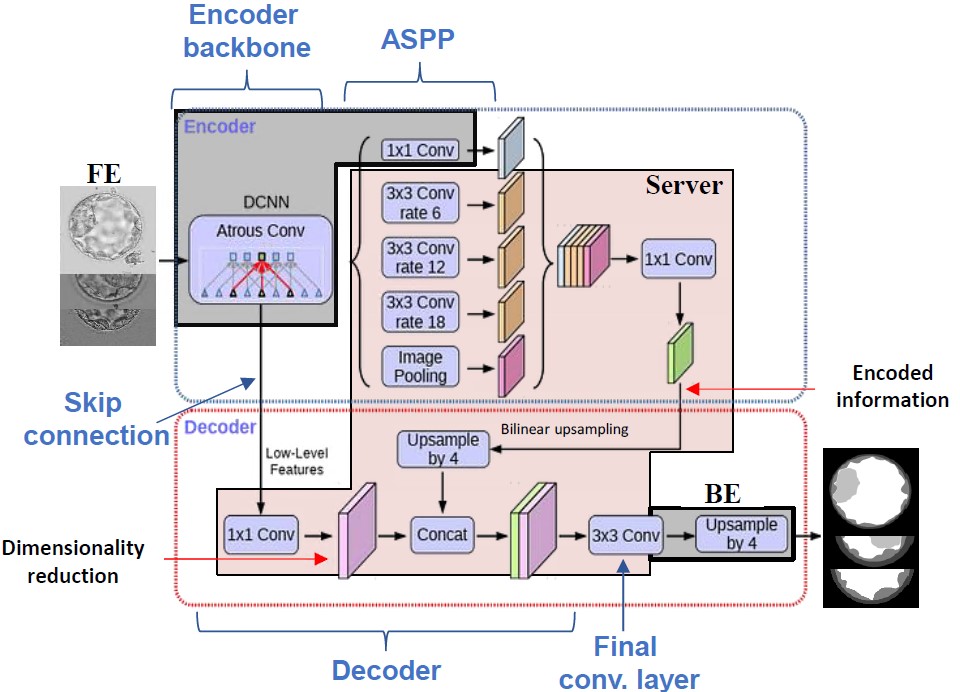}
        \caption{Split DeepLabV3+ architecture with the modified version of Xception as the backbone.}
        \label{fig:DeepLabV3+}
    \end{subfigure}
    \caption{Split DeepLab architectures.}
    \label{fig:combined_deeplab}
\end{figure}

DeepLabV1 addresses those challenges presented by the previous SOTA networks, especially the FCN. It addressed the challenge of reduced feature resolution by employing atrous convolutions for upsampling. It addressed the challenge of reduced localization accuracy, due to DCNN invariance, by performing a post-processing procedure via conditional random fields (CRF)~\cite{chen_2014_Deeplab_V1}. DeepLabV2 addresses the challenge of handling objects at multiple scales by introducing the Atrous Spatial Pyramid Pooling (ASPP) method~\cite{chen_2017_Deeplabv2}. DeepLabV3 represents a significant advancement over DeepLabV2. Unlike DeepLabV2, which relies on the VGG16 backbone and a simpler ASPP module, DeepLabV3 introduces flexibility with diverse backbones like ResNet\cite{He_2016} and Xception\cite{chollet2017xception}. The ASPP module is enhanced with parallel atrous convolutions, capturing multi-scale information more effectively. DeepLabV3 also incorporates global context through image-level features, distinguishing it from its predecessor. The subsequent DeepLabV3+ introduces additional efficiency with depth-wise separable convolution. In summary, DeepLabV3 refines segmentation through advanced techniques, offering improved performance and adaptability over DeepLabV2\cite{deeplabV3}. Fig.~\ref{fig:DeepLabV3} shows the split version of DeepLabV3 with ResNet50 as the backbone used in our work. DeepLabV3+ refines the segmentation output even further, especially along object boundaries. This is achieved using atrous separable convolutions in the encoder-decoder architecture and modifying the backbone version of the Xception network\cite{chollet_2017_xception}. Fig.~\ref{fig:DeepLabV3+} shows the split version of the DeepLabV3+ used in this work.

\subsubsection{RefineNet}
RefineNet is a generic multi-path refinement network that explicitly exploits all the information available along the down-sampling process to enable high-resolution prediction using long-range residual connections~\cite{lin_2017_refinenet}. The authors highlighted limitations associated with typical CNNs, FCNs, and dilated convolutions. CNNs suffer from the downscaling of the feature maps, while FCNs output low-resolution predictions. Dilated convolutions are computationally expensive to train and could quickly reach memory limits. In RefineNet, fine-grained features from earlier convolutions are used directly to improve the deeper layers to capture high-level semantic features. This is called multi-path refinement. Chained residual pooling is also introduced by this work, where rich background context are captured in an efficient manner. Some variants of RefineNet are also proposed as single RefineNet, 2-cascaded RefineNet, and 4-cascaded 2-scale RefineNet that upgrade its overall flexibility. Fig.~\ref{fig:RefineNet} shows the split version of RefineNet used in this work.

\subsubsection{SUNet}
\begin{figure}[t]
    \centering
    \begin{subfigure}[b]{0.47\textwidth}
        \centering
        \includegraphics[width=\textwidth]{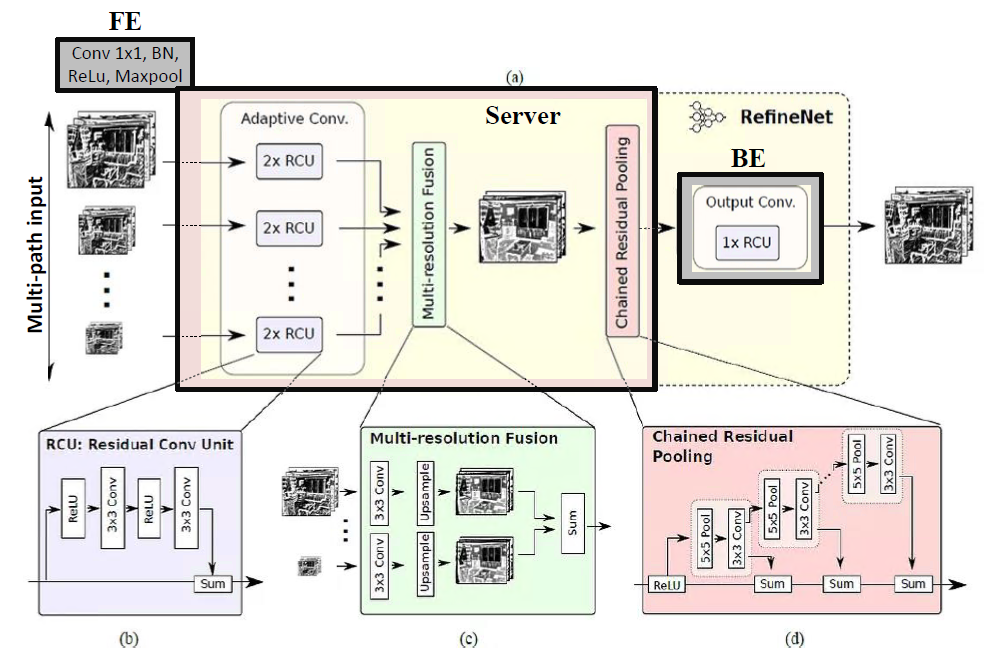}
        \caption{Split RefineNet architecture.}
        \label{fig:RefineNet}
    \end{subfigure}
    \hfill
    \begin{subfigure}[b]{0.50\textwidth}
        \centering
        \includegraphics[width=\textwidth]{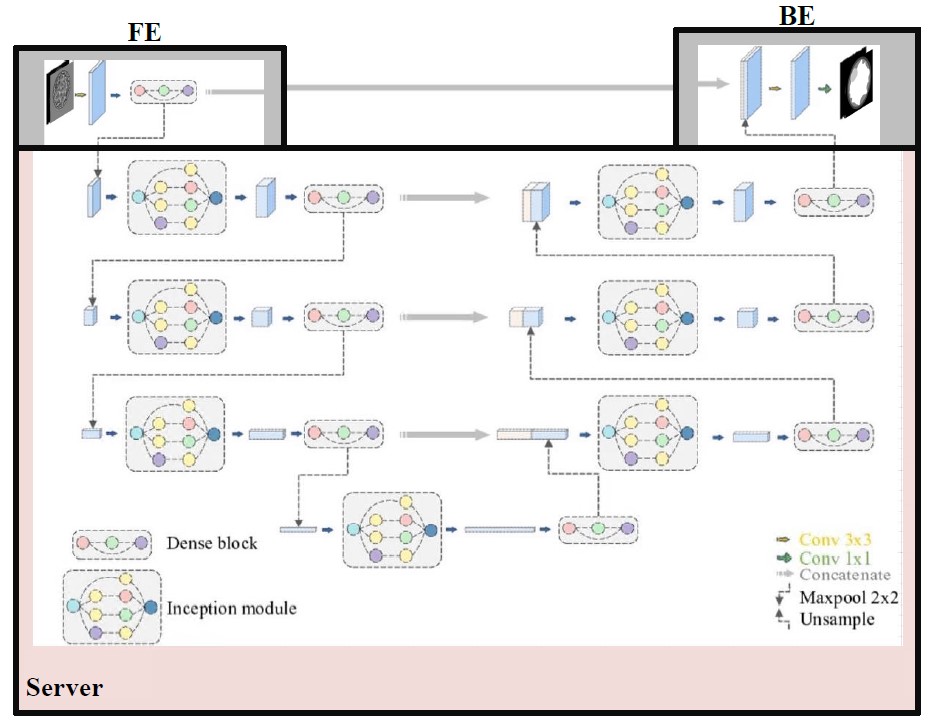}
        \caption{Split SUNet architecture, including inception-dense modules in both encoder and decoder.}
        \label{fig:sunet}
    \end{subfigure}
    \caption{Split RefineNet and SUNet architectures.}
    \label{fig:combined_refinenet_sunet}
\end{figure}

SUNet\cite{SU_Net_Yi_2020} is another U-shaped encoder-decoder network based on the inception module~\cite{inception} and the dense block~\cite{dense_block} to enhance the feature extraction and information reuse capabilities of the network. The idea behind the invention of SUNet is to make the standard UNet stronger in both width and depth. Four versions of SUNet are proposed as SUNet-V1, SUNet-V2, SUNet-V3, and SUNet-V4. As the version number increases, an improvement of classification and segmentation accuracy has been observed. It was originally invented as a federated brain tumor segmentation network. Fig.~\ref{fig:sunet} shows the structure of the split SUNet-V4 architecture. 

\subsubsection{CGNet}
CGNet is a recent network architecture designed for efficient and accurate semantic segmentation\cite{wu_2020_cgnet}. It stands out for having a lightweight architecture that benefits from context information to enhance segmentation performance. CGNet employs context blocks to capture contextual information at different scales, which helps the model to better understand the relationships between different objects and parts of an image. To increase the receptive field without adding to the computational overhead, CGNet utilizes dilated convolutions in the context blocks. Dilated convolutions allow the network to incorporate a larger context while keeping the number of parameters and computations low. The CGNet architecture uses feature fusion modules to combine features from different levels of the network. Such a fusion helps in integrating multi-scale information and improving segmentation performance. The network incorporates skip connections to help propagate information across different layers of the network. These connections enable better gradient flow during training and facilitate the overall optimization process. Despite being lightweight, CGNet delivers competitive performance on various semantic segmentation benchmarks, demonstrating its effectiveness in producing accurate segmentation results. Fig.~\ref{fig:cgnet} shows the split version of CGNet used in our work.

\subsubsection{DUCK-Net}
\begin{figure}[t]
    \centering
    \begin{subfigure}[b]{0.48\textwidth}
        \centering
        \includegraphics[width=\textwidth]{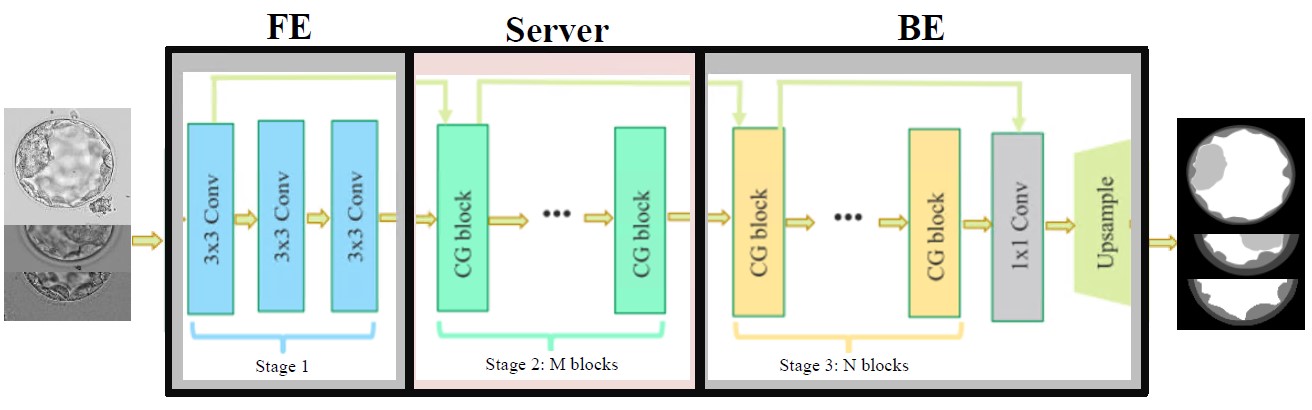}
        \caption{Split CGNet architecture.}
        \label{fig:cgnet}
    \end{subfigure}
    \hfill
    \begin{subfigure}[b]{0.48\textwidth}
        \centering
        \includegraphics[width=\textwidth]{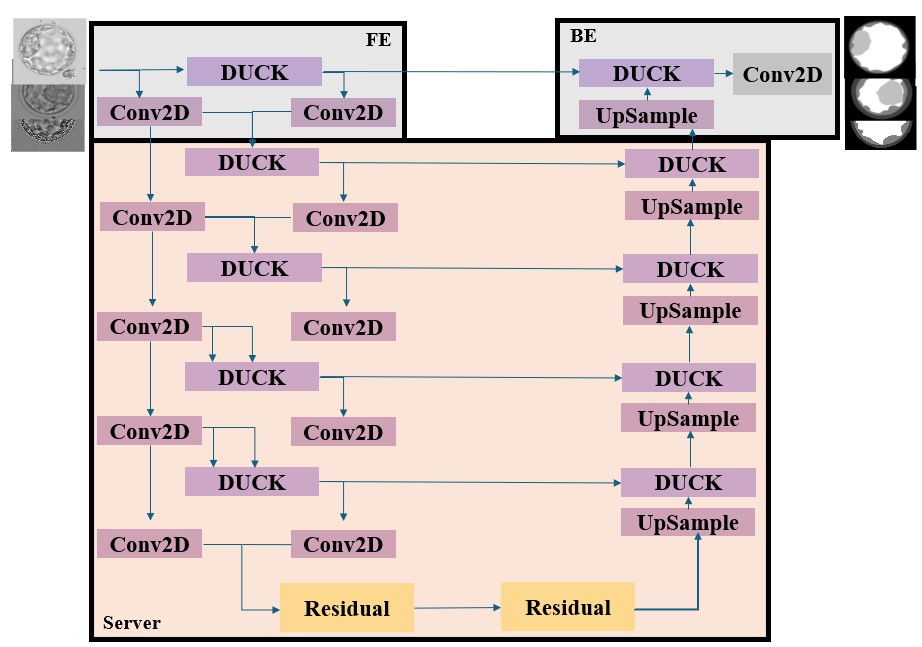}
        \caption{Split version of the DUCK-Net architecture.}
        \label{fig:DUCKNet}
    \end{subfigure}
    \caption{Split CGNet and DUCK-Net architectures.}
    \label{fig:combined_cgnet_ducknet}
\end{figure}

The DUCK-Net~\cite{dumitru2023_ducknet} uses an encoder-decoder architecture and has two main parts: an FCN block called DUCK that utilizes six different types of convolutional blocks at the same time, and a secondary UNet that keeps the low-level information intact. At each step, the DUCK block replaces the traditional pair of 3x3 convolutional blocks used in UNet. This allows the model to capture more details at each step while sacrificing finer, low-level information. Residual blocks are utilized in the last downsampling operation. Fig.~\ref{fig:DUCKNet} shows the split version of DUCK-Net used in this work.

\subsubsection{Attention UNet}
The Attention UNet, a modified UNet architecture for semantic image segmentation\cite{oktay_2018_attentionunet}, integrates attention mechanisms\cite{vaswani_2017_attention} to enhance fine-grained detail capture and precise object localization. This attention mechanism enables the network to focus on informative regions while suppressing less relevant areas.
Following an encoder-decoder structure, Attention UNet extracts hierarchical representations using the encoder and employs attention gates (channel-spatial gates) in the decoder. These gates assign weights to different regions in the feature maps, enhancing the focus on more important details during up-sampling. This attention-driven approach enhances the accuracy of segmentation masks by selectively emphasizing critical features, aiding in precise object and boundary localization. The split version of the Attention UNet used in our work is shown in Fig.~\ref{fig:attunet}. 

\subsubsection{Swin-UNet}
\begin{figure}[t]
    \centering
    \begin{subfigure}[b]{0.48\textwidth}
        \centering
        \includegraphics[width=\textwidth]{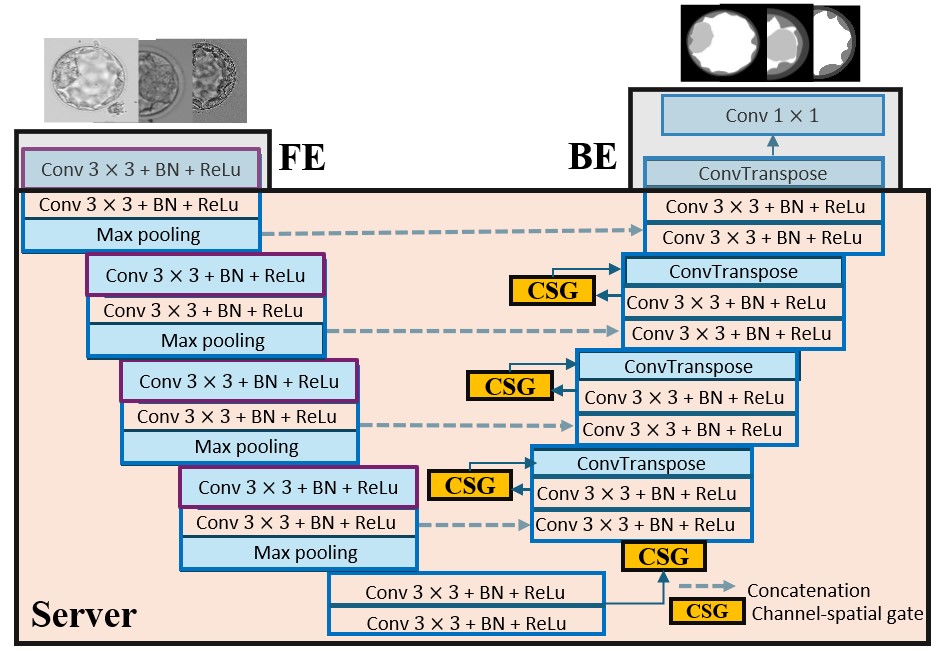}
        \caption{Split Attention UNet architecture.}
        \label{fig:attunet}
    \end{subfigure}
    \hfill
    \begin{subfigure}[b]{0.48\textwidth}
        \centering
        \includegraphics[width=\textwidth]{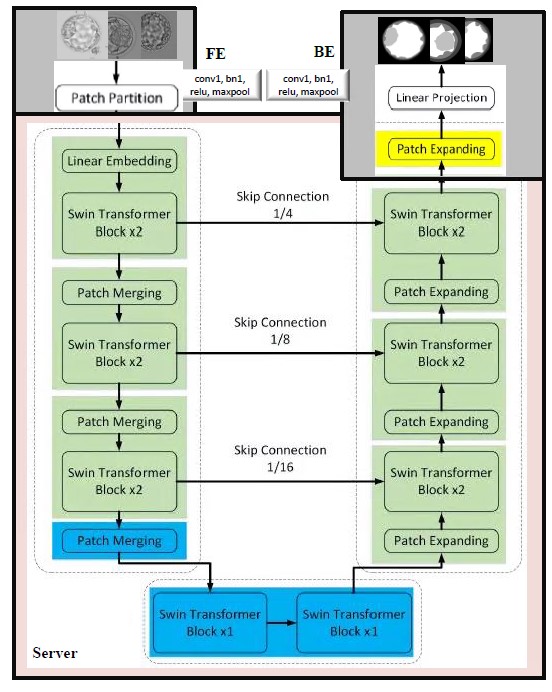}
        \caption{Split Swin-UNet architecture.}
        \label{fig:swinunet}
    \end{subfigure}
    \caption{Split Attention UNet and Swin-UNet architectures.}
    \label{fig:combined_attunet_swinunet}
\end{figure}

Swin-UNet is the first pure transformer-based U-shaped architecture\cite{cao_2022swin}, which combines the architectural advantages of the UNet and the Swin transformer~\cite{liu2021swin}. This network includes the components of patch partition blocks, linear embedding blocks, Swin transformer blocks, patch merging blocks, patch expanding blocks, and linear projection blocks. Three skip connections are used to fuse the multi-scale features from the encoder with the upscaled features. Swin-UNet is famous for comparatively better performance than others due to its efficient attention mechanism, hierarchical feature representation, versatility, fewer parameters, and architectural design. Fig.~\ref{fig:swinunet} shows the split version of the Swin-UNet used in this work.

\subsection{Decision on Split Points Selection}
In a SplitFed network, the selection of the most suitable split points is a crucial task. This choice of \textit{where to split} the model is important to maintain the performance, communication, and overall efficiency of the SplitFed network~\cite{shiranthika2023splitfed}. Our designed U-shaped split networks contain three sub-models: the client-side front-end (FE), the server-side (Server), and the client-side back-end (BE). In this section, we explain the criteria for designing split network architectures through the selection of split points.

\begin{enumerate}
  \item \textbf {Task-specific concerns:} The choice of split points is often guided by the nature of the machine learning task. For instance, in natural language processing, splits should occur at layers that capture semantic features, whereas in computer vision, splits should be at layers that capture high-level visual features. 
  \item \textbf{Communication constraints:} Split points should be  strategically selected to minimize the overall computational load and communication costs associated with information transfer. This involves choosing points where computations are most intensive or sensitive, thus reducing overall latency and communication overhead.
  \item \textbf{Model architecture:} Split points are selected at layers representing high-level features to enable clients to effectively learn task-specific details, ensuring that the model architecture supports the desired learning outcomes. Moreover, the edge blocks maintain the same dimensions, which is necessary for backpropagating gradients in the backward pass. Each sub-model generates its own gradients, making consistent dimensionality crucial.
  \item \textbf{Privacy and security concerns:} To maintain data privacy, splits must be designed so that sensitive data remains on the client side. This approach involves creating two distinct model parts on the client side, with the front end handling sensitive data and the back end managing sensitive GTs. 
  \item \textbf{Computational capabilities of clients:} Split points are chosen to ensure that clients perform minimal computations, allowing those with limited resources to participate in the SplitFed training process without facing computational constraints.
\end{enumerate}

In our experiments, we approached the determination of split points by evaluating each network individually. We carefully considered one or more of the criteria outlined above to ensure that the split points were chosen in a way that optimally addressed the specific characteristics and requirements of each network. This methodical approach allowed us to tailor our decisions to the unique aspects of each network, thereby enhancing the overall effectiveness of our experiments.

\section{Experiments}
\label{Sec3}
\label{sec:experiments}
In this section, we detail our experimental setup and results, followed by an evaluation that includes comparisons of performance and computational complexity.

\subsection{Experimental Setup}
We carried out the experiments with the three medical datasets introduced in Section~\ref{sec:introduction}: 

- \textbf{Blastocyst} dataset: includes 801 Blastocyst RGB images along with their GTs created for a multi-class embryo segmentation task. Each image is segmented into five classes: zona pellucida (ZP), trophectoderm (TE), blastocoel (BL), inner cell mass (ICM), and background (BG).

- \textbf{HAM10K} dataset: Human Against Machine dataset contains 10,015 dermatoscopic RGB images along with the corresponding binary GT masks, representing seven different types of skin lesions, including melanoma and benign conditions. Each image is segmented into two classes: skin lesion and background.

- \textbf{KVASIR-SEG} dataset: contains 1,000 annotated endoscopic RGB images of polyps from colonoscopy procedures, each paired with a binary GT segmentation mask. Each image is segmented into two classes: abnormal condition (such as a lesion, polyp, or ulcer) and background.

We randomly distributed samples from each dataset between clients. For the Blastocyst dataset, samples were assigned to four clients with 110, 90, 200, and 300 samples, respectively, with an additional 101 samples reserved for testing. The HAM10K dataset was divided among ten clients, receiving 1176, 588, 305, 941, 1058, 1294, 648, 942, 883, and 1132 samples each, with 1000 samples set aside for testing. The KVASIR-SEG dataset was partitioned into four clients with 125, 175, 275, and 325 samples, respectively, and 100 images reserved for testing. In each segmentation task, 85\% of the data was used for training and 15\% for validation by each client.

All samples were resized to uniform dimensions of $240 \times 240$ pixels. Based on the network architecture and the dataset, we used several augmentation techniques to improve individual model training performance. Those include horizontal and vertical flipping, rotating, RGB shifting, normalizing, random brightness contrasting, etc. \textit{Soft Dice Loss}~\cite{sudre_2017} was chosen as the loss function. We used the Adam optimizer with network-specific and dataset-specific initial learning rates.  \textit{Intersection Over Union} (IoU) averaged over the sample size (average IoU) or the average Jaccard Index~\cite{Cox_2008} was used as the performance metric. For the Blastocyst dataset, we utilized average IoU of the TE, ZP, BL, and ICM components. Each centralized model was trained for 120 epochs. Each SplitFed model was trained for 10 global communication rounds, while each client trained their local models for 12 local epochs.  

The training process begins with each client receiving randomly initialized copies of the FE and BE sub-models, along with a copy of the server sub-model. Clients then train these models locally for a fixed number of epochs, collaborating with their respective server sub-model copies. After local training, clients send  the weights of their FE and BE sub-models to the server, where they are aggregated to form an updated global model. The updated sub-models are then distributed back to the clients for local validation, completing one global epoch. During training, clients send features from the FE sub-model to the server, which processes them and returns the output for BE processing. The BE sub-model generates predictions, computes the loss, and initiates back-propagation, with gradient updates passed from the BE sub-model to the server and then back to the FE sub-model.

\begin{table*}[ht]
\centering
\renewcommand*{\arraystretch}{1.3}
\caption{Performance Comparison: Centralized (C), Locally Centralized (L), and SplitFed (S) Models (Average IoU metrics)}
\label{tab:merged_comparison}
\setlength{\tabcolsep}{2pt}
\begin{tabular}{|>{\raggedright\arraybackslash}p{75pt}|>
{\raggedright\arraybackslash}p{35pt}|>
{\raggedright\arraybackslash}p{35pt}|>
{\raggedright\arraybackslash}p{35pt}|>
{\raggedright\arraybackslash}p{35pt}|>
{\raggedright\arraybackslash}p{35pt}|>
{\raggedright\arraybackslash}p{35pt}|>
{\raggedright\arraybackslash}p{35pt}|>
{\raggedright\arraybackslash}p{35pt}|>
{\raggedright\arraybackslash}p{35pt}|}
\hline 
\multirow{2}{4em}{\textbf{Model}} & \multicolumn{3}{c|} {\textbf{Blastocyst Dataset}} &
\multicolumn{3}{c|} {\textbf{HAM10K Dataset}} &
\multicolumn{3}{c|}{\textbf{KVASIR-SEG Dataset}} \\
\cline{2-10}
\textbf{} & \textbf{C} & \textbf{L} & \textbf{S} & \textbf{C} & \textbf{L} & \textbf{S} & \textbf{C} & \textbf{L} & \textbf{S} \\ 
\hline 
\textbf{UNet} & 0.8643 & 0.7726 & 0.8593 & 0.8672 & 0.8320 & 0.8640 & 0.8271 & 0.6946 & 0.8042 \\ \hline
\textbf{SegNet} & 0.8475 & 0.7416 & 0.8475 & 0.8426 & 0.7773 & 0.8620 & 0.7337 & 0.5713 & 0.7669 \\ \hline
\textbf{SUNet} & 0.8487 & 0.7566 & 0.8504 & 0.8679 & 0.8241 & 0.8539 & 0.7280 & 0.6006 & 0.7233 \\ \hline
\textbf{DeepLabV3} & 0.8768 & 0.8016 & 0.8369 & 0.8699 & 0.7715 & 0.8696 & 0.8438 & 0.7715 & 0.8262 \\ \hline
\textbf{DeepLabV3+} & 0.8774 & 0.6834 & 0.8591 & 0.8715 & 0.8311 & 0.8262 & 0.8264 & 0.6965 & 0.8278 \\ \hline
\textbf{RefineNet} & 0.7881 & 0.6948 & 0.8181 & 0.8584 & 0.8161 & 0.8403 & 0.7083 & 0.6669 & 0.7652 \\ \hline
\textbf{Attention UNet} & 0.8673 & 0.6990 & 0.8605 & 0.8654 & 0.8241 & 0.8699 & 0.8236 & 0.6991 & 0.7961 \\ \hline
\textbf{Swin-UNet} & 0.8074 & 0.6283 & 0.8142 & 0.8492 & 0.7768 & 0.8478 & 0.7871 & 0.5483 & 0.6642 \\ \hline
\textbf{CGNet} & 0.8433 & 0.7287 & 0.7891 & 0.8728 & 0.8382 & 0.8490 & 0.8354 & 0.6868 & 0.8110 \\ \hline
\textbf{DUCK-Net} & 0.8725 & 0.7994 & 0.8321 & 0.8652 & 0.8389 & 0.8600 & 0.8824 & 0.7778 & 0.7800 \\ \hline
\textbf{Average over models} &  0.8493 & 0.7306 & 0.8367 & 0.8630 & 0.8130 & 0.8543 & 0.7996 & 0.6713 & 0.7765 \\ \hline
\end{tabular}
\end{table*}

\subsection{Experimental Results}
The following sections outline our experimental results.
We show the results in two sections: Quantitative results and Qualitative results.

\subsubsection{Quantitative Results}
\label{sec:exp_quantitative}
We consider cases:

\begin{itemize}
\item  \textbf{Centralized learning on full data:}
We initially trained each network without splitting for image segmentation. We utilized the entire data from the three datasets separately. The average IoUs for all data samples in each set for the centralized models are displayed in the \textbf{C} column of Table~\ref{tab:merged_comparison}.

\item \textbf{Centralized learning locally at each client:}
Secondly, we trained each client's local data in a client-specific, centralized manner to ensure a fair comparison. In this step, each client trained the networks without data splitting. We recorded the IoUs for each client and computed the average, which is presented in the \textbf{L} column for each segmentation task in Table~\ref{tab:merged_comparison}.

\item \textbf{SplitFed learning:}
Thirdly, we trained the SplitFed networks in collaboration with all clients. The IoUs of the SplitFed models are recorded in the \textbf{S} column for each segmentation task in Table~\ref{tab:merged_comparison}.
\end{itemize}

\subsubsection{Qualitative Results}
We included a visual comparison of three random samples from each test set during the SplitFed model training, as shown in Table~\ref{tab:qualitative_comparison}.


\begin{table*}
\centering
\renewcommand*{\arraystretch}{1.5}
\caption{Qualitative comparison of three samples from SplitFed training}
\label{tab:qualitative_comparison}
\setlength{\tabcolsep}{1.5pt}
\begin{tabular}{|>{\raggedright\arraybackslash}p{38pt}|>
{\raggedright\arraybackslash}p{40pt}|>
{\raggedright\arraybackslash}p{40pt}|>
{\raggedright\arraybackslash}p{40pt}|>
{\raggedright\arraybackslash}p{40pt}|>
{\raggedright\arraybackslash}p{40pt}|>
{\raggedright\arraybackslash}p{40pt}|>
{\raggedright\arraybackslash}p{40pt}|>
{\raggedright\arraybackslash}p{40pt}|>
{\raggedright\arraybackslash}p{40pt}|>
{\raggedright\arraybackslash}p{40pt}|>
{\raggedright\arraybackslash}p{40pt}|}
\hline 
\textbf{Sample} & \textbf{Ground Truth} & \textbf{UNet}  & \textbf{SegNet}  &  \textbf{SUNet}    
& \textbf{DeepLab V3}  & \textbf{DeepLab V3+}  &  \textbf{RefineNet}
& \textbf{Attention UNet}  & \textbf{Swin-UNet}  &  \textbf{CGNet} &  \textbf{DUCK-Net}

\\
 \hline 
\multicolumn{12}{|c|}{\textbf{Blastocyst Dataset}} \\ \hline
\parbox[c]{0.55in}{%
    \vspace{-0.25in}%
    \begin{center}\includegraphics[width=0.55in,height=0.55in]{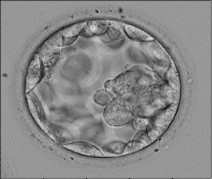}\\
        {\tiny Blast\_PCRM\_R14-0411a.BMP}
    \end{center}
}
&
{\centerline{\includegraphics[width=0.55in,height=0.55in]{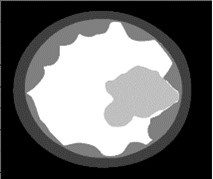}}} & {\centerline{\includegraphics[width=0.55in,height=0.55in]{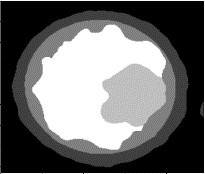}}} & 

{\centerline{\includegraphics[width=0.55in, height=0.55in]{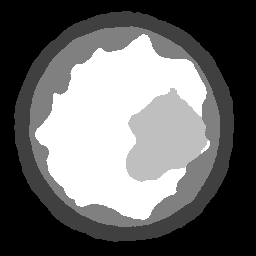}}}
& 
{\centerline{\includegraphics[width=0.55in, height=0.55in]{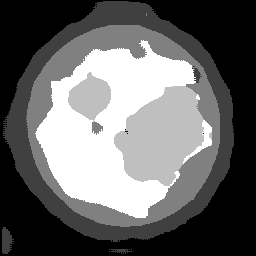}}}
&
{\centerline{\includegraphics[width=0.55in, height=0.55in]{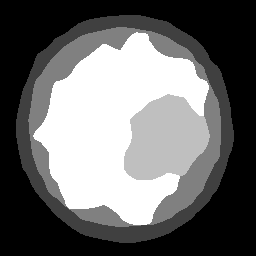}}}
& 
{\centerline{\includegraphics[width=0.55in,height=0.55in]{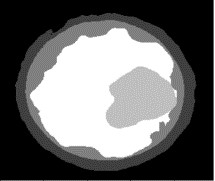}}}&
{\centerline{\includegraphics[width=0.55in,height=0.55in]{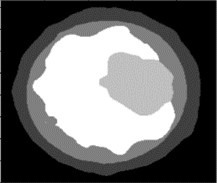}}}& 
{\centerline{\includegraphics[width=0.55in, height=0.55in]{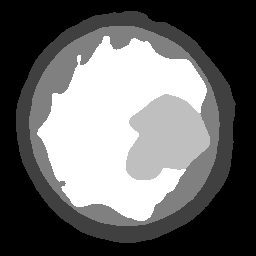}}}
&
{\centerline{\includegraphics[width=0.55in,height=0.55in]{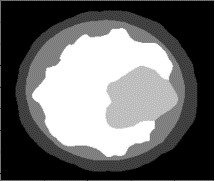}}}&
{\centerline{\includegraphics[width=0.55in,height=0.55in]{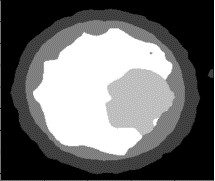}}}& 
{\centerline{\includegraphics[width=0.55in, height=0.55in]{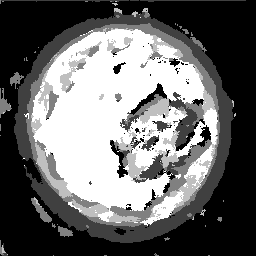}}}\\ \hline

\multicolumn{12}{|c|}{\textbf{HAM10K Dataset}} \\ \hline
\parbox[c]{0.55in}{%
    \vspace{-0.35in}%
    \begin{center}\includegraphics[width=0.55in,height=0.55in]{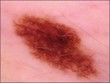}\\
        {\tiny ISIC\_0024308.jpg}
    \end{center}
}
&
{\centerline{\includegraphics[width=0.55in, height = 0.55in]{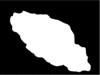}}} & {\centerline{\includegraphics[width=0.55in, height = 0.55in]{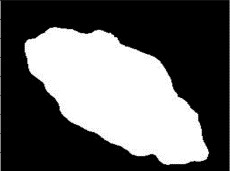}}} & 

{\centerline{\includegraphics[width=0.55in, height=0.55in]{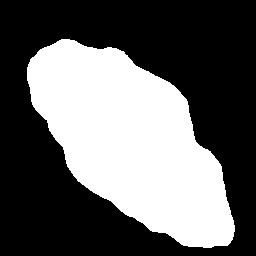}}}
&
{\centerline{\includegraphics[width=0.55in, height=0.55in]{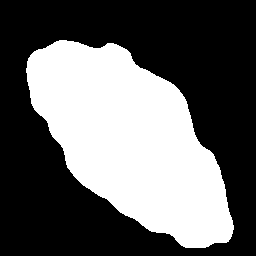}}}

& 
{\centerline{\includegraphics[width=0.55in, height=0.55in]{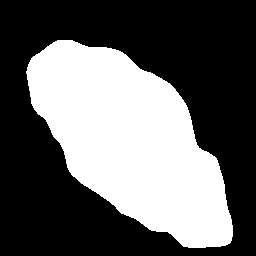}}}

&
{\centerline{\includegraphics[width=0.55in, height=0.55in]{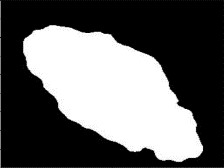}}}&
{\centerline{\includegraphics[width=0.55in, height=0.55in]{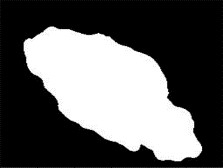}}}&
{\centerline{\includegraphics[width=0.55in, height=0.55in]{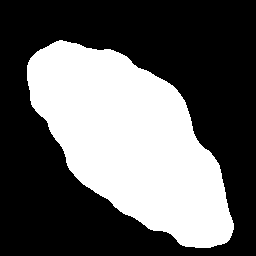}}}

& 
{\centerline{\includegraphics[width=0.55in, height=0.55in]{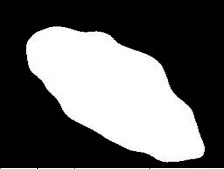}}}&
{\centerline{\includegraphics[width=0.55in, height=0.55in]{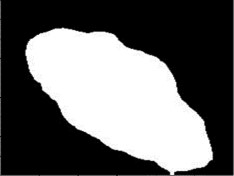}}}&
{\centerline{\includegraphics[width=0.55in, height=0.55in]{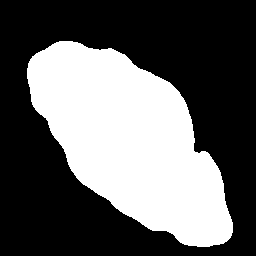}}}
\\ \hline

\multicolumn{12}{|c|}{\textbf{KVASIR-SEG Dataset}} \\ \hline 
\parbox[c]{0.55in}{%
    \vspace{-0.27in}%
    \begin{center}\includegraphics[width=0.55in,height=0.55in]{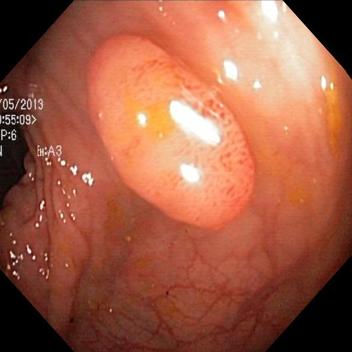}\\
        {\tiny cju7bgnvb1sf8087\\ 17qa799ir.jpg}
    \end{center}
}
&
{\centerline{\includegraphics[width=0.55in]{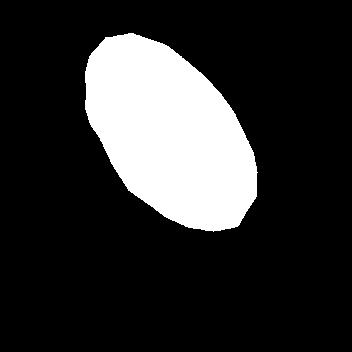}}} & {\centerline{\includegraphics[width=0.55in]{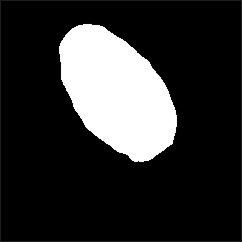}}} & 
{\centerline{\includegraphics[width=0.55in]{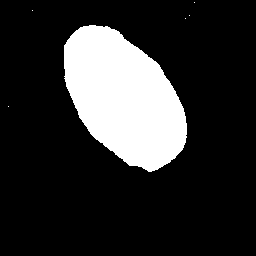}}} 
& 
{\centerline{\includegraphics[width=0.55in]{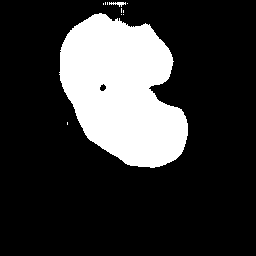}}} 
& {\centerline{\includegraphics[width=0.55in]{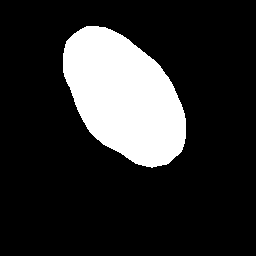}}}
&
{\centerline{\includegraphics[width=0.55in]{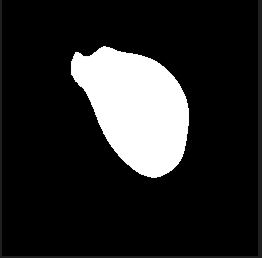}}}&
{\centerline{\includegraphics[width=0.55in]{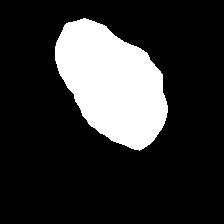}}}& 
{\centerline{\includegraphics[width=0.55in]{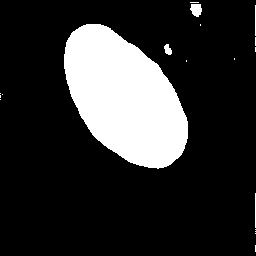}}} 
& 
{\centerline{\includegraphics[width=0.55in]{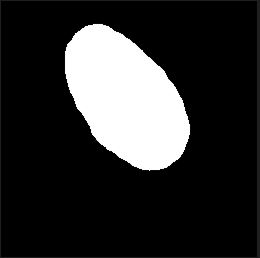}}}&
{\centerline{\includegraphics[width=0.55in]{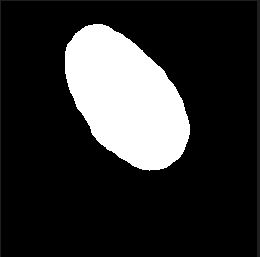}}}& {\centerline{\includegraphics[width=0.55in]{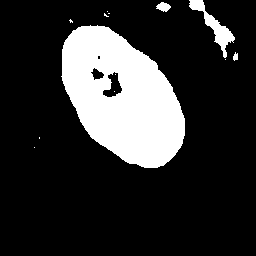}}}
\\ \hline

\end{tabular}
\end{table*}

\subsection{Evaluation}
\subsubsection{Testing Performance Comparison}
\label{sec:evaluation}
When evaluating the cases described in Section~\ref{sec:exp_quantitative}, we expect the following IoU behaviors
: Centralized learning on full data (\textbf{C}) should achieve the highest average IoUs due to access to the complete dataset, enabling better generalization. SplitFed learning should follow with slightly lower average IoUs, as collaboration among clients is effective but may face communication and synchronization challenges. Centralized learning locally (\textbf{L}) at each client is expected to yield the lowest average IoUs, as limited data at each client reduces generalization and segmentation performance. Table~\ref{tab:merged_comparison} confirms this: the highest average IoUs (of individual models) and the highest average over models are in the \textbf{C} column, followed by the \textbf{S} column, with the lowest in the \textbf{L} column. When looking at the average over the models (last row), the \textbf{C} column in the Blastocyst dataset showed 10.61\% higher average IoU than the \textbf{L} column, and 1.26\% higher average IoU than the \textbf{S} column. \textbf{C} column in the HAM10K dataset showed 5\% higher average IoU than the \textbf{L} column, and 0.87\% higher average IoU than the \textbf{S} column. \textbf{C} column in the KVASIR-SEG dataset showed 12.83\% higher average IoU than the \textbf{L} column, and 2.01\% higher average IoU than the \textbf{S} column. 

Table~\ref{tab:testingiou_comparison} shows the three top-performing models for each case of the centralized, locally centralized, and SplitFed for the three segmentation tasks. As seen from the results, for smaller datasets like the Blastocyst dataset, models like Attention UNet and SUNet perform better during SplitFed, likely due to their ability to better generalize with limited data. Complex models like DeepLabV3 still perform well in both centralized and SplitFed, due to their pattern recognition capabilities. In the HAM10K dataset, which is much larger, more complex models such as CGNet and DeepLabV3 excel in the centralized case because they can better utilize the complete dataset for complex feature extraction. Even in SplitFed, models like DeepLabV3 and UNet continue to perform well, benefiting from efficient feature extraction suited to larger datasets. In the KVASIR-SEG dataset, models like DeepLabV3 and CGNet consistently perform well across all cases, showing their versatility. DUCK-Net also performs strongly in centralized cases, likely due to its efficiency in handling polyp segmentation tasks. However, in SplitFed cases, the Duck block at the edge layers appears to rely on tight synchronization, which may result in inefficiencies and slower convergence. 

\begin{figure}[t]
\centerline{\includegraphics[scale = 0.47]{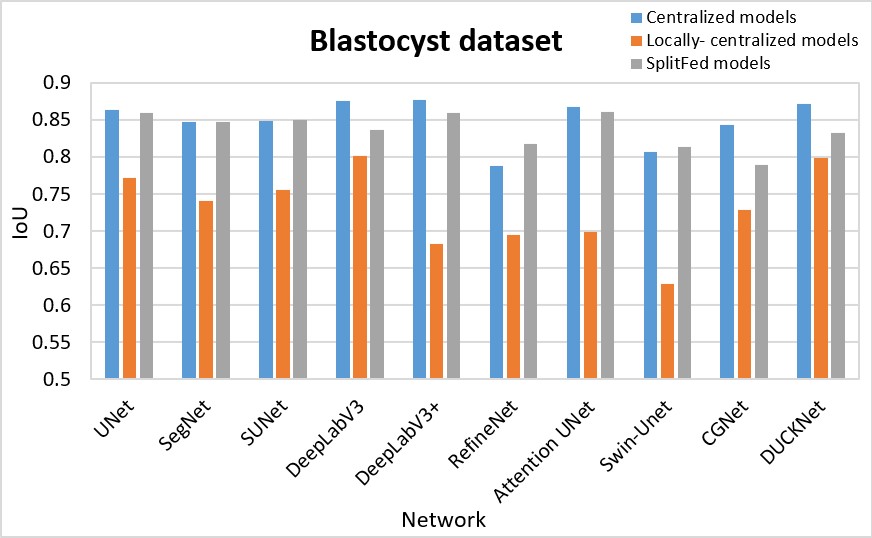}}
\caption{Comparison of results: Blastocyst Dataset. The blue-colored segment is applied to the centralized models; the orange-colored segment is applied to the locally centralized models; and the ash-colored segment is applied to the SplitFed models. Note that the same convention is applied in Fig.~\ref{fig:HAM10KALL} and Fig.~\ref{fig:KVASIRSEGALL}.}
\label{fig:BlastoALL}
\end{figure}

\begin{figure}[t]
\centerline{\includegraphics[scale = 0.47]{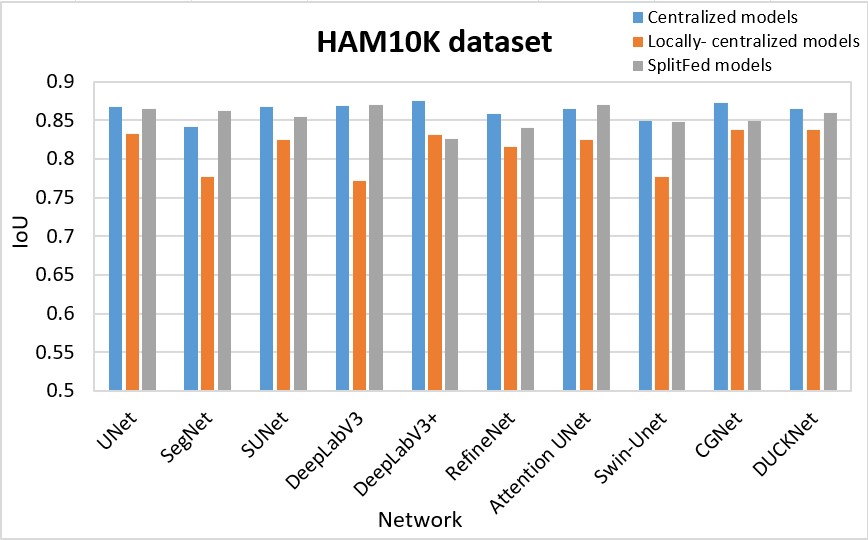}}
\caption{Comparison of results: HAM10K Dataset.}
\label{fig:HAM10KALL}
\end{figure}

\begin{figure}[t]
\centerline{\includegraphics[scale = 0.47]{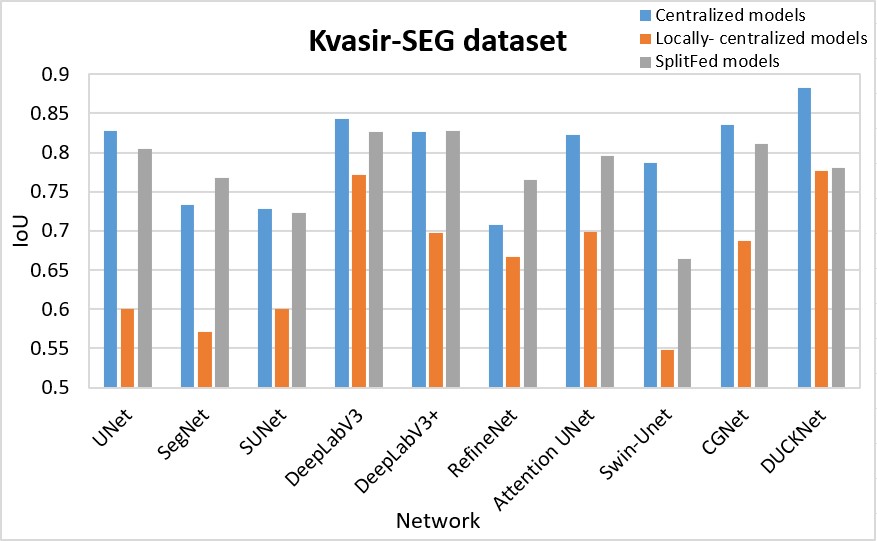}}
\caption{Comparison of results: KVASIR-SEG Dataset.}
\label{fig:KVASIRSEGALL}
\end{figure}

\begin{table}
\centering
\LTcapwidth=\textwidth
\renewcommand*{\arraystretch}{1.0}
\caption{Performance-wise comparison for SplitFed networks} (Three top performing networks in each category are listed)
\label{tab:testingiou_comparison}
\setlength{\tabcolsep}{3pt}
\begin{tabular}{|>{\raggedright\arraybackslash}p{75pt}|>
{\raggedright\arraybackslash}p{80pt}|>
{\raggedright\arraybackslash}p{80pt}|>
{\raggedright\arraybackslash}p{80pt}|}
\hline 
\textbf{Model}  & \textbf{Centralized models}  & \textbf{Locally centralized models} & \textbf{SplitFed models}  \\
 \hline 
\textbf{Blastocyst dataset} & DeepLabV3+, DeepLabV3, DUCK-Net  & DeepLabV3, DUCK-Net, CGNet & Attention UNet, DeepLabV3+, SUNet \\ \hline
\textbf{HAM10K dataset} & CGNet, DeepLabV3+, DeepLabV3 & DUCK-Net, CGNet, DeepLabV3+ & Attention UNet, DeepLabV3, UNet  \\ \hline
\textbf{KVASIR-SEG dataset} & DUCK-Net, DeepLabV3, CGNet & DUCK-Net, DeepLabV3, DeepLabV3+ & DeepLabV3+, DeepLabV3, CGNet  \\ \hline
\end{tabular}
\end{table}
For the purpose of better comparison, we graphically represent the results in Table~\ref{tab:merged_comparison} in Figs.~\ref{fig:BlastoALL},~\ref{fig:HAM10KALL}, and~\ref{fig:KVASIRSEGALL}. Fig.~\ref{fig:BlastoALL} summarizes the segmentation performance of various methods for the Blastocyst dataset, while Fig.~\ref{fig:HAM10KALL} shows the results for the HAM10K dataset. Comparisons for the KVASIR-SEG dataset are displayed in Fig.~\ref{fig:KVASIRSEGALL}. In these figures, the performance of the centralized model is represented in blue, the performance of the locally centralized models is depicted in orange, and the performance of the SplitFed models is illustrated in gray. From these figures, the SplitFed models have demonstrated superior performance compared to the locally-centralized models, with performances closer to their corresponding centralized models. 

\subsubsection{Comparison of Computational Complexity} 
We measured each model's complexity with floating-point operations (FLOPs) by using the \textit{ptflops} package in Python~\cite{ptflops}. Table~\ref{tab:comcomplex_comparison} outlines the number of trainable parameters of each network, the measured FLOPs, the number of trainable parameters, and the FLOPs value with respect to UNet as the anchor. According to these metrics, CGNet has the least number of trainable parameters and the lowest FLOPs values compared to the other networks. RefineNet has the highest percentage of trainable parameters, while Attention UNet has the highest FLOPs value. 

\begin{table}
\centering
\LTcapwidth=\textwidth
\renewcommand*{\arraystretch}{1.0}
\caption{Computational complexity-wise comparison} (FLOPs are measured in 'MAC', which stands for 'Multiply-Accumulate')
\label{tab:comcomplex_comparison}
\setlength{\tabcolsep}{3pt}
\begin{tabular}{|>{\raggedright\arraybackslash}p{65pt}|>
{\raggedright\arraybackslash}p{60pt}|>
{\raggedright\arraybackslash}p{60pt}|>
{\raggedright\arraybackslash}p{60pt}|>
{\raggedright\arraybackslash}p{60pt}|}
\hline 
\textbf{Model}  & \textbf{Trainable parameters (TP)}  & \textbf{FLOPs}  &  \textbf{TP as a \%. of UNet} & \textbf{FLOPs as a \%. of UNet} \\
 \hline 
\textbf{UNet} & 7.76M & 10.52 GMAC & 1\% & 1\% \\ \hline
\textbf{SegNet} & {9.44 M} & {7.04 GMAC} & {1.22\%} &0.67\% \\ \hline
\textbf{SUNet} & {14.1 M} & {24 GMAC} & {1.82\%} & 2.28\% \\ \hline
\textbf{DeepLabV3} & {28.32 M } & {12.83 GMAC} & {3.64\%} & 1.21\% \\ \hline
\textbf{DeepLabV3+} & 54.70 M  & 15.85 GMAC & 7.05\% & 1.50\% \\ \hline
\textbf{RefineNet} & 118 M & 50.24 GMAC & 15.20\% & 4.77\% \\ \hline
\textbf{Attention UNet} & 34.87 M & 51.03 GMAC & 4.50\% & 4.85 \%\\ \hline
\textbf{Swin-UNet} & 41.38 M & 8.67 GMAC & 5.33\% & 0.82\%\\ \hline
\textbf{CGNet} & 0.30 M & 541.69 MMAC & 0.039\% &0.05\%\\ \hline
\textbf{DUCK-Net} & {22.67 M }  & {12.55 GMAC}  & {2.92\%} & 1.19\%  \\ \hline

\end{tabular}
\end{table}
\subsubsection{Performance Comparison with Other Existing Methods}
Table~\ref{tab:SOTA_comparison_All} presents the average IoUs of the other existing methods for semantic segmentation with the Blastocyst, HAM10K, and KVASIR-SEG datasets. Please note that we only included those methods that clearly reported their average IoUs or average Jaccard Indices for the purpose of a fair comparison. Studies that solely mentioned accuracy metrics or any other metrics were intentionally omitted from the evaluation.

\begin{table}
\centering
\LTcapwidth=\textwidth
\renewcommand*{\arraystretch}{1.3}
\caption{Performance comparison with other existing methods} (NA: Not Applicable/ Relevant experimental results are not shown/experiments have not been conducted.)
\label{tab:SOTA_comparison_All}
\setlength{\tabcolsep}{2pt}
\begin{tabular}{|>{\raggedright\arraybackslash}p{85pt}|>
{\raggedright\arraybackslash}p{65pt}|>
{\raggedright\arraybackslash}p{65pt}|>
{\raggedright\arraybackslash}p{65pt}|
}
\hline 
\textbf{SOTA research}  & \textbf{Centralized models IoU}  & \textbf{Federated models IoU} & \textbf{SplitFed models IoU}   \\
 \hline 

\multicolumn{4}{|c|}{\textbf{Blastocyst Dataset}} \\ \hline
Our Previous research&  0.798 (BLAST-NET-\cite{Rad_2019}), 0.817~\cite{shiranthika2023splitfed} & 0.810~\cite{shiranthika2023splitfed} &0.825~\cite{shiranthika2023splitfed}   \\ \hline

\multicolumn{4}{|c|}{\textbf{HAM10K Dataset}} \\ \hline
FedPerl (Efficient-Net)\cite{bdair2021fedperl} & 0.769 & 0.747 & NA \\ \hline
FedMix (UNet)\cite{wicaksana2022fedmix} & NA & 0.819 $\pm$ 1.7 & NA\\ \hline
MALUNet\cite{ruan2022malunet} & 0.802 & NA &NA \\ \hline
FedZaCt (UNet)\cite{yang2022fedzact} & 0.855 & 0.856 &NA \\ \hline
FedZaCt (DeepLabV3+)\cite{yang2022fedzact} & 0.861 & 0.863 &NA \\ \hline
Chen \etal~\cite{Chen_2022} & NA & 0.892 &NA \\ \hline
FedDTM (UNet)\cite{liu2023medical} & NA & 0.7994 &NA \\ \hline

\multicolumn{4}{|c|}{\textbf{KVASIR-SEG Dataset}} \\ \hline
DUCK-Net\cite{dumitru2023_ducknet} & 0.9502 & NA &NA  \\ \hline
FCN-Transformer\cite{sanderson2022fcn} & 0.9220& NA & NA\\ \hline
MSRF-Net\cite{srivastava2021msrf} & 0.8508 & NA &NA \\ \hline
PraNet\cite{fan2020pranet} & 0.7286 & NA &NA \\ \hline
HRNetV2\cite{HRNet_2019} & 0.8530 & NA &NA \\ \hline
Subedi \etal~\cite{subedi2023client} & 0.81 & 0.823 &NA \\ \hline
DilatedSegNet ~\cite{tomar2023dilatedSegnet} & 0.8957 & NA &NA \\ \hline
DeepLabV3+~\cite{tomar2023dilatedSegnet} & 0.8837 & NA &NA \\ \hline
Colonformer~\cite{duc2022colonformer} & 0.877 & 0.876 &NA \\ \hline
SSFormer-S~\cite{wang2022stepwise} & 0.8743 & NA &NA \\ \hline
SSFormer-L~\cite{wang2022stepwise} & 0.8905 & NA &NA \\ \hline
FedDM~\cite{zhu2023feddm} & 0.5275$\pm$0.0002 & 0.6877$\pm$0.0308 &NA \\ \hline
\end{tabular}
\end{table}

\section{Future Directions Associated with Decentralized Learning Network Architectures}
\label{future_works}
Throughout our work presented here, we identified several intriguing research directions that align with the network repository we developed. These are outlined in this section.

\subsection{Transformers in SplitFed}
The integration of transformer architectures into decentralized learning has recently gained attention, demonstrating exceptional performance in medical segmentation. Their success stems from features like attention mechanisms, hierarchical representation, scalability, and interpretability, which allow them to capture spatial relationships and complex patterns in medical images, outperforming CNN-based models~\cite{schlemper2019attention, wang2018non, chen2021transunet, xiao2023transformers}. Our results with Attention UNet and Swin-UNet further support these advantages, highlighting the promising future of transformers in semantic segmentation tasks.

\subsection{Design of Cross-Domain Architectures}
Cross-domain analysis is well-established in ML and DL architectures~\cite{Alimova_2022, khetani2023cross}, but its exploration in SplitFed or FL remains in its early stages. In ML and DL, cross-domain analysis aids in identifying algorithmic strengths and limitations during knowledge transfer, accelerating solution development while reducing time and effort. In the medical field, dynamic and subjective data distributions present a significant challenge. This opens opportunities for innovation, particularly in designing semantic segmentation models that can generalize across diverse medical domains or adapt to varying client data.

\subsection{Personalized Network Architectures}
Designing semantic segmentation networks that dynamically adjust complexity, depth, and architectural parameters based on each client's device or data capabilities is a promising direction. This aligns with the concept of ``personalized federated learning" in FL~\cite{Tan_2023, kulkarni2020survey}. Customized architectures tailored to individual data distributions can enhance global model accuracy, reliability, and generalizability. Such adaptations would result in more effective predictions and recommendations, ultimately improving the global model’s performance across diverse healthcare participants.

\subsection{Privacy-Preserving Architectures}
In decentralized training, it would be beneficial to design neural architectures that embed secure computing methods, such as homomorphic encryption or differential privacy, directly into the network structures. In our research, we support this direction by developing a neural architecture repository for SplitFed networks. This repository lays the groundwork for integrating these privacy-preserving techniques into collaborative learning models, ensuring data security while maintaining the efficiency of SplitFed architectures. 

\subsection{Communication-Efficient Architectures}
Our experiments revealed significant overhead in data transmission within SplitFed networks. Thus, designing network architectures that minimize data transmission at model split points in collaborative learning is a promising area of research. Techniques such as quantization, pruning, and compression can be integrated to effectively reduce communication overhead.

\subsection{Design of SplitFed-Specific Network Architectures}
It is important to recognize the need for network architectures specifically designed for SplitFed, rather than relying on generic semantic segmentation models. While architecture-focused designs have been explored in the FL context~\cite{lo2022architectural, lo2021flra}, a significant gap remains in the SplitFed domain.

\section{Conclusions}\label{Conclusions}
In this work, we introduced MedSegNet10, a new publicly accessible repository for medical image segmentation utilizing the SplitFed learning mechanism. Our main goal was to create a resource allowing researchers to easily integrate SplitFed models into their own applications without extensive customization by allowing reusing the networks in the MedSegNet10 repository. We designed, implemented, and optimized SplitFed versions of ten selected models to simplify the process and make the strengths of SplitFed accessible to a wider audience. In addition, we outlined several future research directions in medical image segmentation that could benefit from SplitFed learning. As SplitFed continues to evolve, we believe that MedSegNet10 would serve as a valuable foundation for ongoing developments, guiding researchers toward more precise and impactful solutions in medical image segmentation.

\section*{Acknowledgments}
The authors thank the Natural Sciences and Engineering Research Council (NSERC) of Canada for their financial support. 

\bibliographystyle{IEEEbib}
\bibliography{ref}

\end{document}